\documentclass[notitlepage,floatfix,amsmath,amssymb,pre,twocolumn]{revtex4-1} 
\usepackage{tabularx}
\usepackage{graphicx}
\usepackage{dcolumn}
\usepackage{color}
\usepackage[normalem]{ulem}
\usepackage{amssymb}
\usepackage[toc,page]{appendix}

\newcommand{\be}{\begin{equation}}
\newcommand{\ee}{\end{equation}}

\usepackage{graphicx}
\usepackage{enumerate}
\newcommand{\RN}[1]{%
  \textup{\uppercase\expandafter{\romannumeral#1}}%
}

\begin{document}
\title{Extraction of invariant manifolds and application to turbulence with a passive scalar}

\author{N.E.~Sujovolsky and P.D.~Mininni}
\affiliation{
  Universidad de Buenos Aires, Facultad de Ciencias Exactas y
  Naturales, Departamento de F\'\i sica, \& IFIBA, CONICET, Ciudad
  Universitaria, Buenos Aires 1428, Argentina.}

\begin{abstract}
The reduction of dimensionality of physical systems, specially in
fluid dynamics, leads in many situations to nonlinear ordinary
differential equations which have global invariant manifolds with
algebraic expressions containing relevant physical information of the
original system. We present a method to identify such manifolds, 
and we apply it to a reduced model for the Lagrangian evolution of field
gradients in homogeneous and isotropic turbulence with a passive
scalar.
\end{abstract}
\maketitle

Fluids, and turbulence in particular, have been a playground
for many applications of the theory of dynamical systems. Methods to
reduce the number of degrees of freedom of a system have been
developed for or applied to fluid dynamics, from 
truncations using Fourier \cite{lorenz63} or empirical modes
\cite{holmesberkooz96}, to models that reduce the
dimensionality of the Lagrangian evolution of field gradients
\cite{li_origin_2005, chevillard2006lagrangian,
  meneveau2011lagrangian, Johnson_2016, Pereira_2018, 
  sujovolsky2019invariant}. These latter models often also display
global invariant manifolds (GIMs, i.e., manifolds globally preserved
under the system evolution) with algebraic expressions, as the
Vieillefosse manifold \cite{vieillefosse_local_1982}. Such manifolds
play a crucial role in the dynamics, and perturbative methods were
developed \cite{gucken83} to identify them in nonlinear ordinary
differential equations (ODEs). Here we present a method to identify
GIMs with algebraic expressions, in ODEs as those resulting from
reduced models for fluids, and apply it to homogeneous and isotropic
turbulence (HIT) with a passive scalar \cite{Celani_2000,
  watanabe_2007, buaria_2021}. Examples of passive scalars include
small temperature fluctuations in a fluid, atmospheric humidity, or
chemicals concentration. The concentration also corresponds to the
continuous limit of diluted particles, of interest for practical and
theoretical reasons \cite{Shraiman_2000, Falkovich_2001, Donzis_2010,
  Gotoh_2015}. However, reduced models for passive scalar gradients
have been barely explored. We thus also derive a reduced model for
this case. In HIT, similar models, sometimes 
  called restricted Euler models or QR-models, are used to study
  singularities \cite{vieillefosse_local_1982} and the origin of
  non-Gaussian statistics and intermittency \cite{li_origin_2005},
  among many other applications. Our model in turn can help understand
  how turbulence affects the passive scalar transport.
The results we present are then twofold: A
reduced model for field gradients evolution in passive scalar
turbulence, and a method to identify GIMs.

We consider the Lagrangian evolution of the gradients of an
incompressible velocity field $\bf{u}$ (with $\nabla \cdot {\bf
  u}=0$) and of a passive scalar $\theta$. The dynamics are given by
\begin{equation}
\label{eq:n-s}
  D_t {\bf u} = -{\boldsymbol \nabla}p + \nu \nabla^2 {\bf u} +
    {\bf f}  , \,\,\,\,\,\,\, 
  D_t \theta = \kappa \nabla^{2} \theta + f_{\theta},
\end{equation}
where $D_t = \partial_t + {\bf u}\cdot{\boldsymbol
  \nabla}$ is the Lagrangian derivative, $p$ the pressure per unit
mass density, $\nu$ the kinematic viscosity, $\kappa$ the diffusivity,
${\bf f}$ an external mechanical forcing, and $f_{\theta}$ a scalar
source. For $\nu = \kappa$ the equations have one dimensionless
parameter that controls the dynamics, the Reynolds number
$\mathrm{Re}=u \lambda/\nu$, where $u$ is the r.m.s.~flow velocity
and $\lambda$ the Taylor microscale, a characteristic scale in
turbulence.
To compare with the reduced model we performed a direct numerical
simulation (DNS) of Eq.~(\ref{eq:n-s}) in a three-dimensional (3D)
periodic cubic domain with spatial resolution of $1024^{3}$ grid
points. We used a parallel pseudospectral fully-dealiased method to
compute spatial derivatives and nonlinear terms, and
a second-order Runge-Kutta scheme for time integration
\cite{mininni2011hybrid}. A 3D large-scale random forcing ${\bf f}$ was
used to sustain the turbulence, and a random scalar source
$f_{\theta}$ was applied to inject passive scalar concentration. The
viscosity and diffusivity where chosen in such a way that all the
relevant flow scales were properly resolved. This resulted in
${\textrm Re} \approx 560$, with $k \eta \approx 1.2$, where
$k$ is the maximum resolved wavenumber and $\eta$ the
Kolmogorov dissipation scale. To study the Lagrangian evolution (i.e.,
following fluid elements) of velocity and passive scalar gradients, we
tracked $10^6$ Lagrangian particles for several large-scale
turnover times using the methods discussed in \cite{yeung1988algorithm}.
Velocity and passive scalar gradients at particles' positions
${\bf x}$, ${\boldsymbol \nabla} {\bf u}({\bf x}, t)$ and
${\boldsymbol \nabla} \theta({\bf x}, t)$, were stored with high time
cadence to compute their statistics and time derivatives.

{\it A reduced model for the Lagrangian evolution of field gradients.}
From Eq.~(\ref{eq:n-s}) we now derive a closed model for velocity and
passive scalar field gradients, based on similar models for velocity
gradients in HIT \cite{meneveau2011lagrangian}. 
While HIT is sustained out of equilibrium by the external forcing and
dissipation, we neglect both and consider the ideal unforced case
($\nu = \kappa = {\bf f} = f_{\theta} = 0$). We can thus expect the
reduced model to give a good approximation to field gradients dynamics
only for short times, when the effect of the forcing and of
dissipation are small compared with nonlinearities. In spite of this,
we will see that the reduced model compares well with results from the
DNS. The limit of negligible diffusivity can be interpreted,
  e.g., as the case of passive scalars in atmospheric turbulence, in
  which advection dominates. Diffusivity and viscosity (to also
  consider, e.g., the effect of the Schmidt number) can be later added
  using the methods described in \cite{meneveau2011lagrangian}. 

When computing spatial derivatives of Eq.~(\ref{eq:n-s}), we write
field gradients using index notation and define
$A_{ij}=\partial_{j}u_{i}$ and $\theta_{j}=\partial_{j}\theta$ (for
$i$, $j \in \{x,y,z\}$). Then,
\begin{equation}
\label{eq:Aij} 
	D_t A_{ij} + A_{kj}A_{ik} = - \partial_{ij}p , \,\,\,\,\,\,\,
  D_t \theta_{j} + A_{kj} \theta_{k} = 0,
\end{equation}
where $\partial_{ij} = \partial_{i} \partial_{j}$. The first equation is the
usual Lagrangian evolution equation for the velocity gradient tensor
$A_{ij}$, while the second is the Lagrangian evolution equation for
passive scalar gradients. Some derivatives of the pressure in
Eq.~(\ref{eq:Aij}) can be removed using the incompressibility
condition  ${\boldsymbol \nabla} \cdot {\bf u} = A_{ii}=0$, which for
$D_t A_{ii}$ in Eq.~(\ref{eq:Aij}) implies $A_{ki}A_{ik} = -
\partial_{ii} p$. The remaining spatial derivatives of the pressure
can be written using the deviatoric part of the pressure Hessian,
$H_{ij} = -( \partial_{ij} p - \delta_{ij} \partial_{kk} p /
3)$, where $\delta_{ij}$ is the Kronecker delta. The equation for $D_t A_{ij}$ can then be written as
\begin{equation}
  D_t A_{ij} + A_{kj}A_{ik} -\delta_{ij} A_{kl}A_{lk}/3 =  H_{ij}.
\label{eq:AijH0}
\end{equation}
This equation, together with the equation for $D_t \theta_j$ in
Eq.~(\ref{eq:Aij}), provide a set of equations (albeit not closed) for
the evolution of all components of ${\boldsymbol \nabla}{\bf u}$ and
${\boldsymbol \nabla} \theta$ along the trajectories of fluid
elements.

To close this set of equations we use an approximation commonly used
in restricted Euler models of HIT \cite{cantwell_exact_1992,
  chevillard2006lagrangian, meneveau2011lagrangian}, and assume that
the deviatoric part of the pressure Hessian can be neglected. In other
words, we assume that  $H_{ij}\approx 0$ in
Eq.~(\ref{eq:AijH0}). Attempts have been made to improve this
approximation, which otherwise results in a finite time blow up of the
reduced model for HIT \cite{Wilczek_2014, Carbone_2020,
  Parashar_2020}; such attempts include a multi-scale model that regularizes the
  singularity and retains the geometrical properties of the QR-model
  \cite{Chevillard_2008}. Then we reduce the information in $A_{ij}$
and $\theta_j$ to the smallest possible number of scalar quantities
resulting in an autonomous system. For the velocity gradients in
HIT, two scalar quantities (proportional to the traces of ${\bf A}^2$
and ${\bf A}^3$ and thus invariant under the group of rotations and
reflections, with ${\bf A}$ the velocity gradient tensor) suffice to
obtain the QR-model \cite{meneveau2011lagrangian}. Here, we also want
to take into account the scalar gradients in the
model. Thus we define
\begin{eqnarray}
Q&=&-A_{ij}A_{ji}/2, \ \ R=-A_{ij}A_{jk}A_{ki}/3, \ \ S=  \sum_{i}
  \theta_{i}/3, \nonumber \\
T&=& \sum_{i} \theta_{j}A_{ji}/3, \ \ U = \sum_{i}
  \theta_{j}A_{jk}A_{ki} / 3, 
\label{eq:variables}
\end{eqnarray}
where repeated indices are summed. The equations for the evolution of
$Q$ and $R$ are well known for HIT, and here we briefly outline their
derivation for completeness. To derive an evolution equation for $Q$
we evaluate Eq.~(\ref{eq:AijH0}) in $A_{nj}$ and multiply it by
$A_{in}$, to obtain 
\begin{equation}
D_t (A_{in}A_{nj}) + 2 A_{ik}A_{kn}A_{nj} - \dfrac{2}{3}A_{ij}
A_{kl}A_{lk} = 0,
\label{eq:AinAnj}
\end{equation}  
where $H_{ij}$ was neglected. Setting $i=j$ we obtain
$D_{t}{Q}=-3R$. To obtain an equation for $R$ we multiply
Eq.~(\ref{eq:AinAnj}) by the velocity gradient tensor again, to obtain
 \begin{equation}
D_t(A_{in}A_{nk}A_{kj}) + 3 A_{im}A_{mn}A_{nk}A_{kj} -
  A_{in}A_{nj}A_{kl}A_{lk} = 0.
\label{eq:AinAnkAkj}
\end{equation}
The trace of this equation results in an equation for $R$, but in
order to do so we need to simplify the term
$A_{im}A_{mn}A_{nk}A_{ki}$. We use the Cayley-Hamilton theorem
\cite{vieillefosse_local_1982, cantwell_exact_1992,
  meneveau2011lagrangian}, which for incompressible flows ($A_{ii}=0$)
can be written as $A_{im}A_{mn}A_{nj}=-QA_{ij}-R \delta_{ij}$. Then,
the first term on the r.h.s.~of the trace of Eq.~(\ref{eq:AinAnkAkj}) 
can be written as $A_{im}A_{mn}A_{nk}A_{ki} = - QA_{ik}A_{ki}-R
\delta_{ik}A_{ki}=2Q^{2}$. We finally obtain $D_{t}{R}=2Q^{2}/3$. The
equations for $S$, $T$, and $U$ are new. An equation for $S$ is
obtained by summing over $j$ in the equation for $D_t \theta_j$ in
Eq.~(\ref{eq:Aij}),  resulting in $D_{t}{S}=-T$. An equation for $T$
requires using both $D_t \theta_j$ in Eq.~(\ref{eq:Aij}) multiplied by
$A_{ji}$, and the $ji$-component of Eq.~(\ref{eq:AijH0}) multiplied by
$\theta_j$, to obtain
\begin{equation}
D_t (\theta_{j}A_{ji}) + 2 \theta_{j}A_{jk}A_{ki} - \theta_{i}
  A_{kl}A_{lk}/3  = 0,
\label{eq:T}
\end{equation}
which when summing over $i$ reduces to $D_{t}{T}=-2U - 2SQ/3
$. Finally, to derive an equation for $U$ we use $D_t \theta_j$ in
Eq.~(\ref{eq:Aij}) and the $ji$-component of Eq.~(\ref{eq:AinAnj}) to
write 
\begin{equation}
D_t (\theta_{j}A_{jk}A_{ki}) + 3 \theta_{j}A_{jk}A_{kl}A_{li} -
  2 \theta_{j} A_{ji} A_{kl}A_{lk} /3 = 0 .
\label{eq:Rtheta}
\end{equation}
Using again the Cayley-Hamilton theorem and summing over $i$ results
in $D_{t}U= 5QT/3 + 3RS$. The resulting reduced model for the
Lagrangian evolution of field gradients can be summarized as
\begin{eqnarray}
D_{t}{Q} &=& -3R , \,\,\,\,\,
  D_{t}{R}=2Q^{2}/3, \,\,\,\,\,
  D_{t}{S}=-T, \nonumber \\
D_{t}{T} &=& -2 U - 2SQ/3, \,\,\,\,\,
  D_{t}{U}=-5 QT/3  + 3RS.
\label{eq:ODEs}
\end{eqnarray}
This system prescribes the evolution of field gradients along fluid
elements trajectories. Over these trajectories, the equations are five
closed ODEs. An analogous system can be obtained by considering $i$
to be either $x$, $y$, or $z$ (the same component in $S$, $T$, and
$U$), instead of summing over the three components. This can be useful
in experiments when only one component of the scalar gradients may be
accesible, or for anisotropic flows.

{\it Method to find algebraic GIMs.} Invariant manifolds of a
dynamical system $\dot{\bf y} = {\bf F}({\bf y})$ are manifolds in 
phase space such that initial conditions in the manifold are preserved
inside the manifold by the system evolution. If such manifold can be 
explicitly written as some function $G({\bf y})=c$ with $c$ a
constant, then $\dot{G}({\bf  y})=0$. Moreover, an invariant manifold
in the vicinity of a fixed point $\dot{\bf y} = 0$ with null real
eigenvalues is a central manifold \cite{gucken83}. Such manifolds play
a crucial role as they allow further reductions in the system
dimensionality, and as points in phase space may converge to that
manifold with slow evolution. When these manifolds have global
algebraic expressions, we propose the following method to find them:


{\it (i)} We assign each variable in the ODEs a unit in terms of the
unit of the time derivative, $[D_{t}]=[t^{-1}] = a$. For example, for
$Q$ and $R$ in Eq.~(\ref{eq:ODEs}) (the usual reduced Euler model or
QR-model of HIT), we choose $[Q]=a^{\epsilon_{Q}}$ and
$[R]=a^{\epsilon_{R}}$. We compute the $\epsilon$-exponents for all 
variables (their ``order'') from the ODEs. From the equations for
$D_tQ$ and $D_tR$ in Eq.~(\ref{eq:ODEs}) we get $a^{\epsilon_{Q} + 1}
= a^{\epsilon_{R}}$ and $a^{\epsilon_{R} + 1} = a^{2
  \epsilon_{Q}}$. Their logarithm results in the linear system
\begin{equation}
  \epsilon_{Q} + 1 = \epsilon_{R}, \,\,\,\,\,
  \epsilon_{R} + 1 = 2\epsilon_{Q},
\label{eq:orders}
\end{equation}
and thus $\epsilon_{Q}=2$, $\epsilon_{R}=3$. In some cases the
linear system for the $\epsilon$-exponents can be indeterminated, in
which case information of units from the original physical system,
or the smallest possible exponents, can be used. For the rest of the
variables in Eq.~(\ref{eq:ODEs}), $\epsilon_{S}=1$, $\epsilon_{T}=2$,
and $\epsilon_{U}=3$.

{\it (ii)} We write all algebraic terms of order $n$ in the variables,
for $n \in \mathbb{Z}$ (as from dimensional grounds, all terms in each
GIM must be of the same order). As an example, for the ODEs in
Eq.~(\ref{eq:ODEs}) and for order $n=4$, we have $Q^2$, $Q S^2$,
$Q T$, $R S$, $S^{4}$, $S^{2} T$, $S U$, and $T^{2}$. 

{\it (iii)} To find GIMs, we look for linear combinations of all
terms of order $n$ with a null time derivative. For our ODEs and
$n=4$, the general equation is $D_{t} (k_{1} Q^{2}+ k_{2} QS^{2} +
k_{3} QT + k_{4} R S + k_{5} S^{4} + k_{6} S^{2}T + k_{7} SU + k_{8}
T^{2} ) = 0$. Thus, we compute the $D_{t}$ derivatives of all terms of
order $n$, which give us terms of order $n+1$ (in the example,
$n+1=5$). As an example, $D_{t}( QS^{2}) = -3RS^{2} - 2STQ$.

{\it (iv)} To find linear combinations of $n$th-order terms defining
a GIM, we construct a matrix $\mathbb{C}$ that in the $ij$-cell has
the coefficient multiplying the $i$th-term of order $n+1$ that appears
in the derivative of the $j$th-term of order $n$. In our example, if
we order the columns with the $n=4$ terms as $\{ Q^{2} , Q S^{2}, Q T,
R S,   S^{4}, S^{2} T, S U, T^{2} \}$, and the rows with the $n+1=5$
terms as $\{ QR, RS^{2}, STQ, RT, UQ, SQ^{2}, S^{3}T, ST^{2}, US^{2},
S^{3}Q, TU \}$, then the resulting matrix is
\begin{equation}
	\centering
	\mathbb{C} = 
	\begin{pmatrix}
		-6 & 0 & 0 & 0 & 0 & 0 & 0 & 0 \\
		0 & -3 & 0 & 0 & 0 & 0 & 3 & 0 \\
		0 & -2 & 0 & 0 & 0 & 0 & 5/3 & -4/3 \\
		0 & 0 & -3 & -1 & 0 & 0 & 0 & 0 \\
		0 & 0 & -2 & 0 & 0 & 0 & 0 & 0 \\
		0 & 0 & -2/3 & 2/3 & 0 & 0 & 0 & 0 \\
		0 & 0 & 0 & 0 & -4 & 0 & 0 & 0 \\
		0 & 0 & 0 & 0 & 0 & -2 & 0 & 0 \\
		0 & 0 & 0 & 0 & 0 & -2 & 0 & 0 \\
		0 & 0 & 0 & 0 & 0 & -2/3 & 0 & 0 \\
		0 & 0 & 0 & 0 & 0 & 0 & -1 & -4 \\
	\end{pmatrix}
\label{eq:method_A}
\end{equation}
Note that $\mathbb{C}_{22} = -3$ and $\mathbb{C}_{32} = -2$, as
$D_{t}( QS^{2}) = -3RS^{2} - 2STQ$. The null space of $\mathbb{C}$
gives the generators of the GIMs of order $n$. In this example, the
null space is composed solely by $\left(0,4,0,0,0,0,4,-1 \right)$,
which means that $D_{t}(4QS^{2}+4SU-T^{2})=0$, defining a GIM. To find
all algebraic GIMs of a system, we repeat the process for all
$n \in \mathbb{N}$. The method thus described is reminiscent of
methods used to algebraically equate terms of equal order in the
perturbative search of invariant or central manifolds, and can thus
be also used in such cases to systematically find those manifolds.

\begin{figure}
\includegraphics[width=8cm]{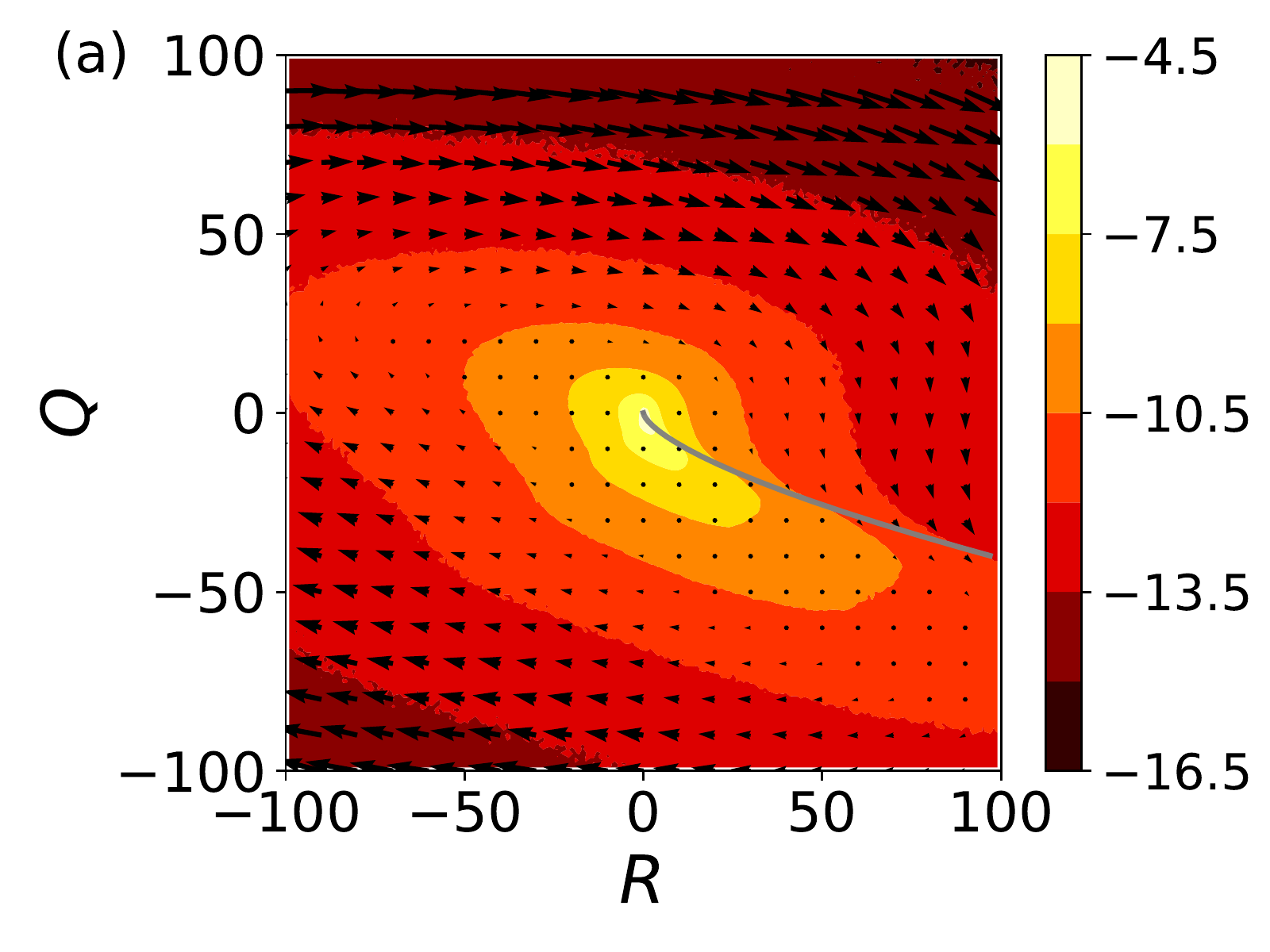}
\includegraphics[width=8cm]{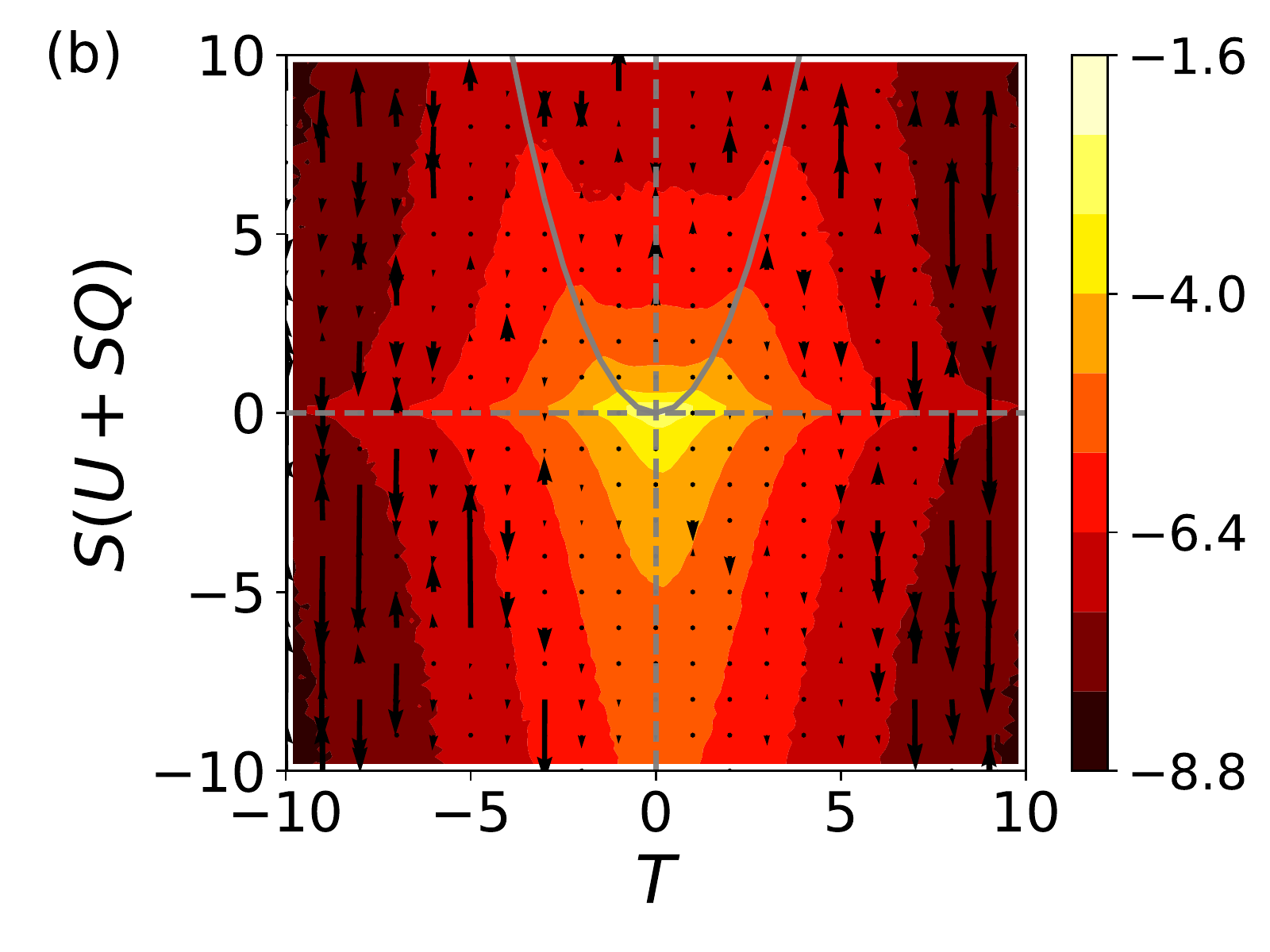}
\caption{(a) Logarithm of the joint PDF of $Q$ and $R$ for fluid
  elements in the DNS (in colors). The Vieillefosse tail $Q=-(27/4
  R^{2})^{1/3}$ is shown by the solid line.
  (b) Same for the joint PDF of $T$ and $S (U+S Q)$. The solid line
  indicates the manifold in Eq.~(\ref{eq:sigma1}), and the dashed
  lines indicates the projection of the fixed points in
  Eq.~(\ref{eq:sigma0}) on this plane. All quantities are
  dimensionless. In both panels the arrows indicate the projection of
  the direction and speed of evolution of fluid elements in the DNS on
  these planes in phase space.}
\label{fig:QR_and_sigma1}
\end{figure}

\begin{figure}
\includegraphics[width=8cm]{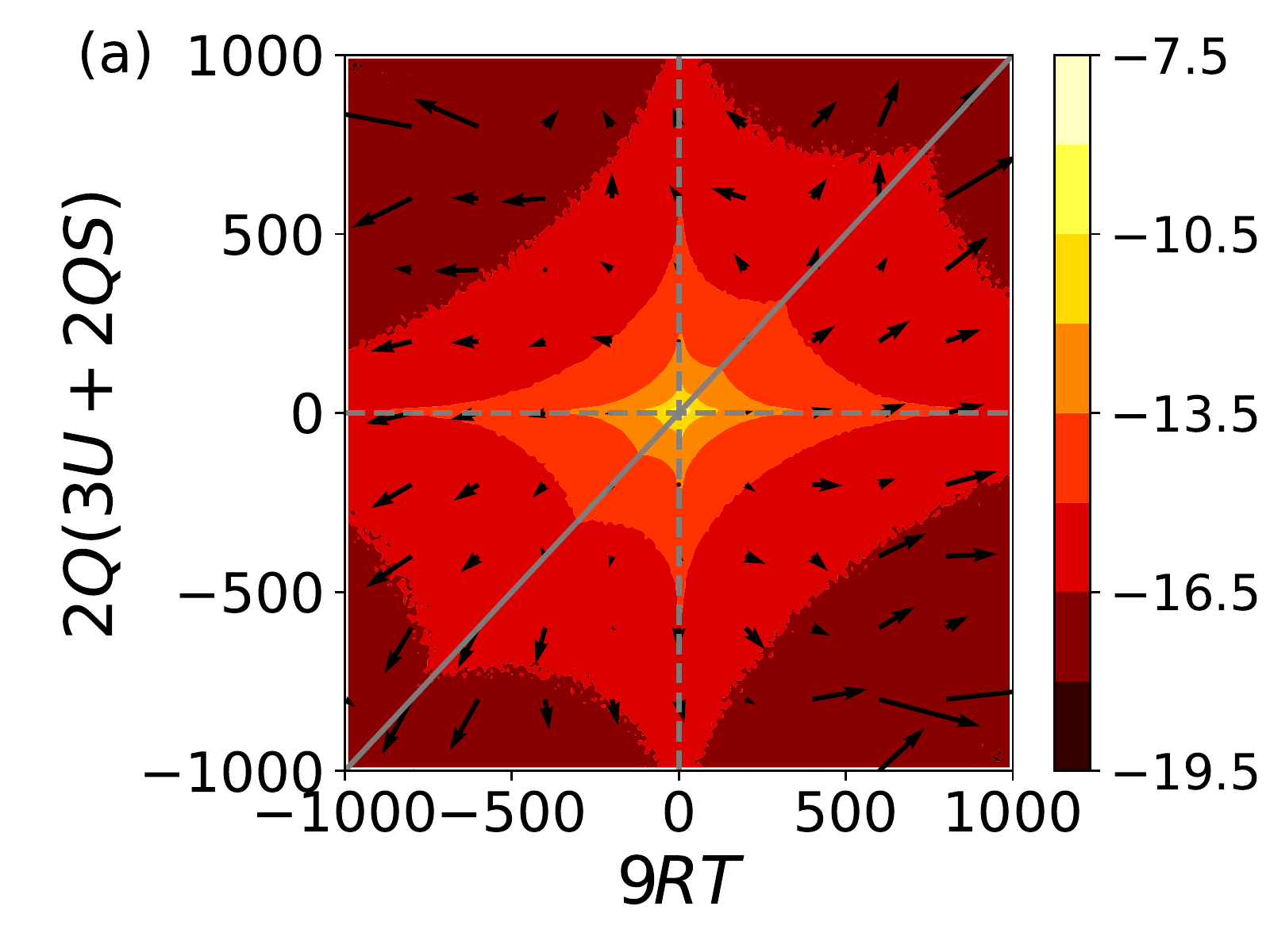}
\includegraphics[width=8cm]{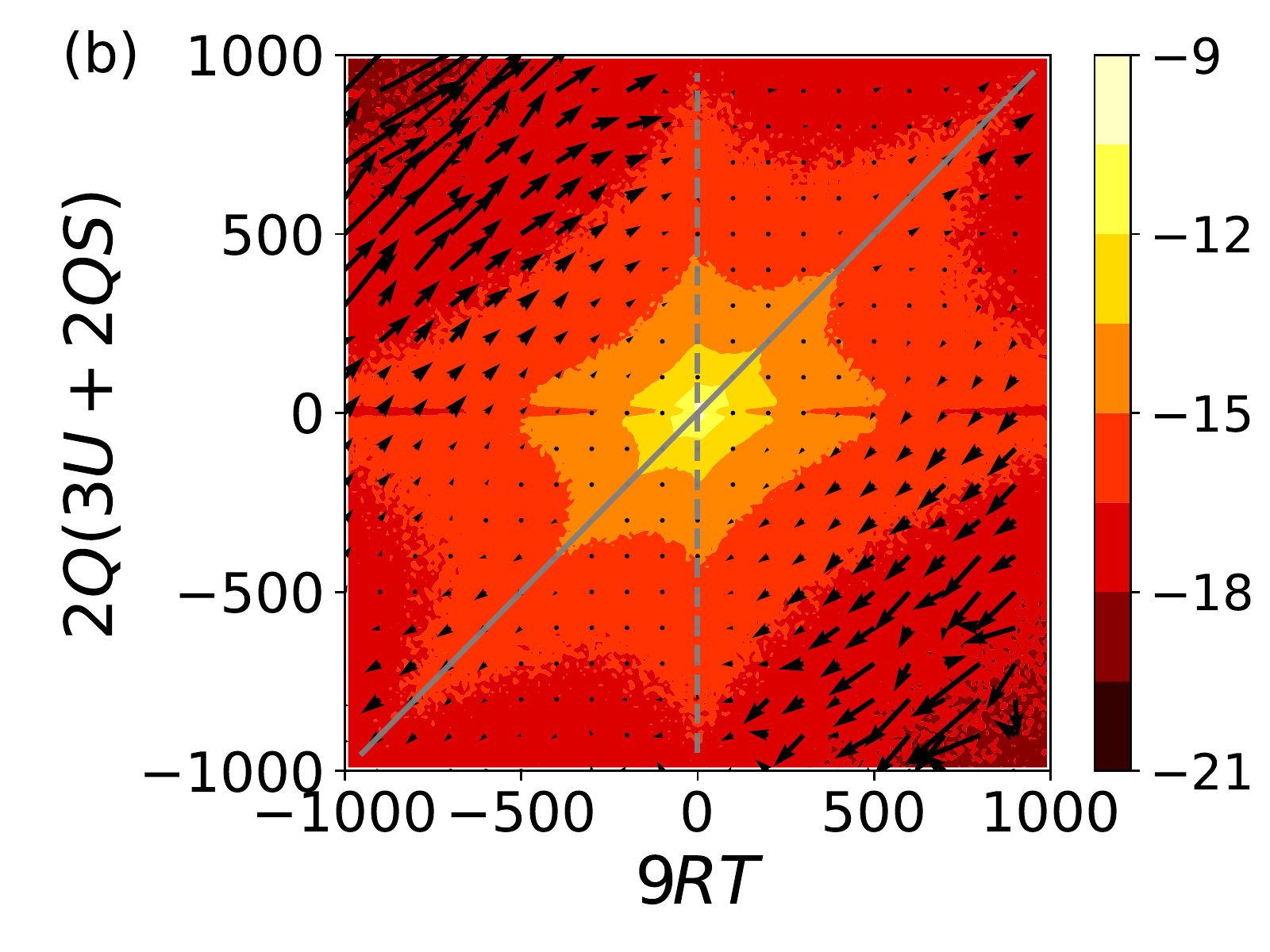}
\caption{Logarithm of the joint PDF (in colors) of $9 R T$ and $2Q
  (3U+2Q S)$ for (a) all fluid elements in the DNS, and (b) fluid
  elements in locations where $|Q|>0.5$, $|R|>0.5$, $|S|>0.5$,
  $|T|>0.5$, and $|U|>0.5$. The manifold in Eq.~(\ref{eq:sigma2}) is
  shown by the solid line, and the projection of the fixed points in
  Eq.~(\ref{eq:sigma0}) on this plane are indicated by the dashed
  lines. References for the arrows are the same as in
  Fig.~\ref{fig:QR_and_sigma1}.}
\label{f:GCIM2}
\end{figure}

{\it Fixed points and GIMs of the reduced Euler model for a passive
  scalar in HIT.} 
The method just described gives the so-called Vieillefosse tail or
invariant manifold of the QR-model of HIT
\cite{vieillefosse_local_1982}, and works with reduced models as those 
obtained for other flows \cite{sujovolsky2019invariant,
  Sujovolsky_2020}. Now we analyze in detail the fixed points and GIMs
of Eq.~(\ref{eq:ODEs}). The Vieillefosse tail
$D_{t}(4Q^{3}+27R^{2})=0$ is obtained as a GIM of Eq.~(\ref{eq:ODEs})
for order $n=6$ (note that while the GIMs of
  the ideal system have free constants of integration, in the presence
  of forcing and disipation they reduce to finite size 
  basins that go through the origin). Figure
\ref{fig:QR_and_sigma1}(a) shows this manifold, as well as the joint
probability density function (PDF) of $Q$ and $R$ for the DNS of the
full set of partial differential equations (with viscosity and
forcing), and also indicates with arrows the evolution in phase space
obtained from the time evolution of 
field gradients along fluid trajectories in the DNS. As is well known
for HIT \cite{meneveau2011lagrangian}, even though the DNS has forcing
and dissipation (and the deviatoric part of the pressure Hessian),
fluid elements have a larger probability of having $Q$ and $R$ close
to the GIM, and also evolve more slowly in the vicinity of this
GIM. This provides valuable information on the geometry of the flow
structures, as the $Q$-$R$ phase space is divided by the Vieillefosse
tail into regions where the flow gradients display different local properties
\cite{cantwell_exact_1992, Dallas_2013}. In spite of the agreements, 
we recall that the QR-model has multiple limitations as, e.g., in the
QR-model trajectories in phase space diverge in finite time, a process
that is arrested in real fluids by pressure gradients and dissipation
\cite{Chevillard_2008, Wilczek_2014, Johnson_2016, Pereira_2018,
  Carbone_2020,Parashar_2020}.

\begin{figure}
\includegraphics[width=8cm]{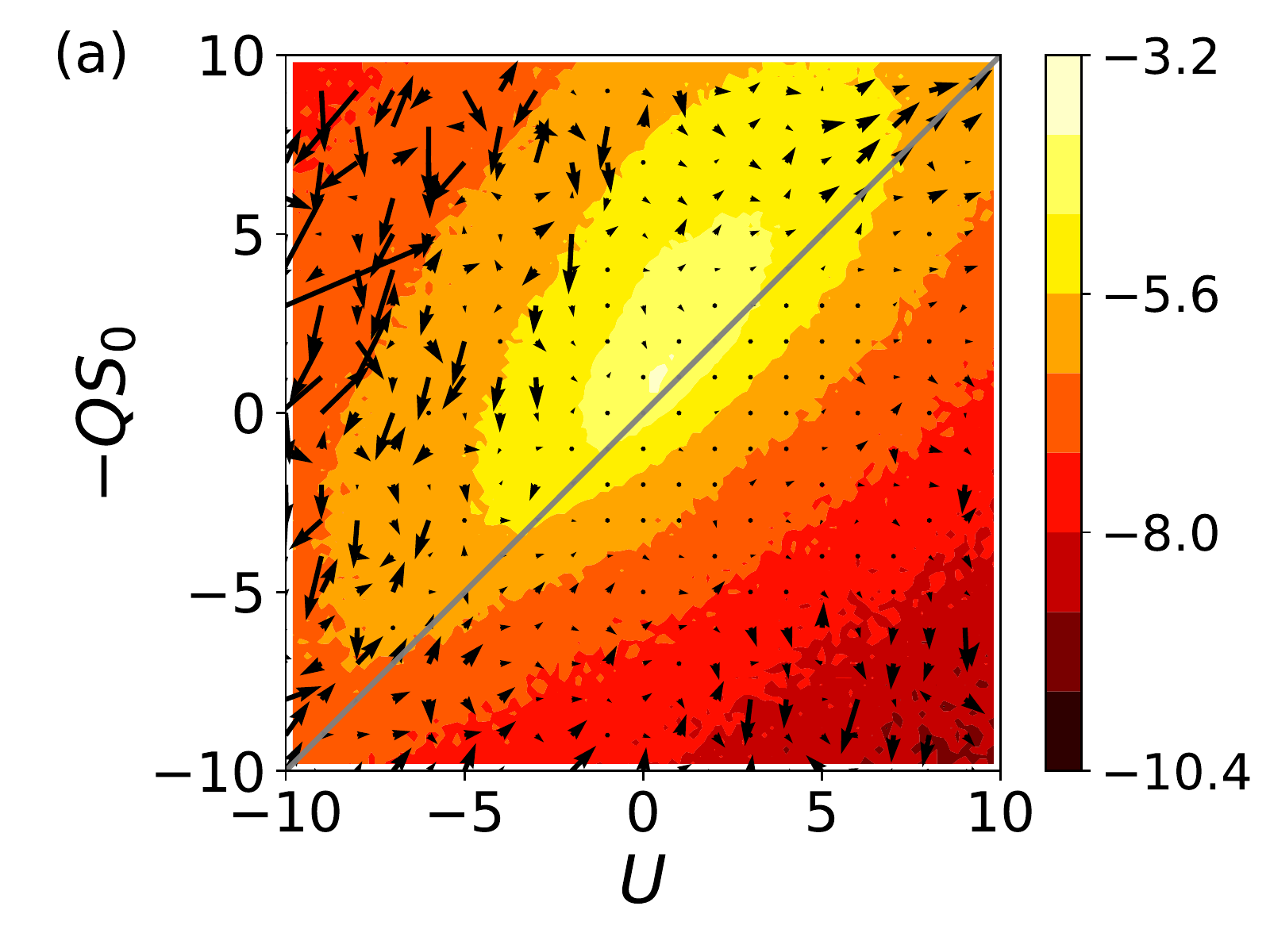}
\includegraphics[width=8cm]{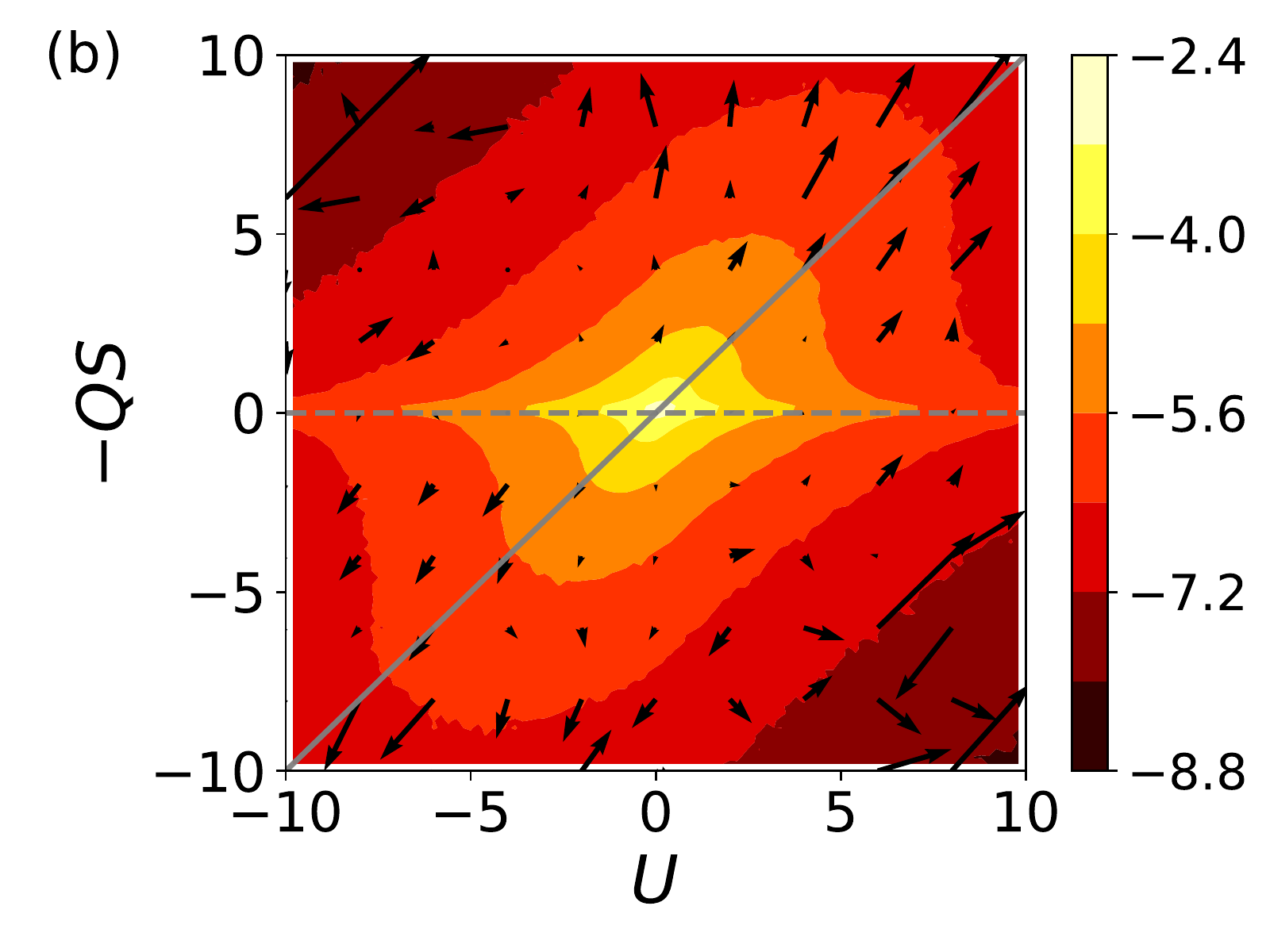}
\caption{Logarithm of the joint PDF of $- Q S$ and $U$ for (a) only
  fluid elements in the DNS with $S \approx S_0 = 1$, and (b) all
  fluid elements. The local invariant manifold $U + Q S_0
  =0$ is shown as a reference by the solid line. In (b) we also 
  show $QS=0$ with dashed lines. Arrows show the same as in
  Figs.~\ref{fig:QR_and_sigma1}.}
\label{f:local_im}
\end{figure}

Equation (\ref{eq:ODEs}) contains the QR-model in its first two
equations, and thus it is not a surprise that the Vieillefosse tail is
a GIM. Let's now consider the fixed points and other
GIMs and local invariant manifolds of the system (mostly related to
the evolution of passive scalar gradients). The fixed
points are a GIM by themselves
\begin{equation}
	\label{eq:sigma0}
	Q=R=T=U=0, \ \ S \ \text{free}.
\end{equation}
For any value of $S=S_0$ these conditions do not evolve in time, and
thus they are an invariant manifold. Physically, they express the
fact that gradients of the scalar field do not change if velocity
field gradients are zero.

Before, we showed that for order $n=4$
\begin{equation}
	\label{eq:sigma1}
	D_{t}(4QS^{2}+4SU-T^{2})=0 ,
\end{equation}
and thus it is a GIM. Figure \ref{fig:QR_and_sigma1}(b) shows the
joint PDF of $S (U+S Q)$ and $T$ in the DNS, indicating
with a solid line $S (U+S Q) = T^2$ (the GIM), as well as $T=0$
and $S (U+S Q) = 0$ (which are projections of the fixed
points on the plane); note the system phase space has 5
dimensions, and thus the two-dimensional PDFs we show are
projections of the total phase space on different planes (this is also
the reason why arrows in the figures may overlap). We again see larger
probabilities of finding fluid elements 
near the manifolds. As before, the arrows show the projection of the
flow on this plane of phase space obtained from the DNS. The GIMs are
not attracting or repelling but arrows become smaller in their
vicinity. Note also how the arrows increase as $T$ increases, in
agreement with equation $D_{t}S=-T$ in Eq.~(\ref{eq:ODEs}): As $T$
increases, so does the variation of $S$. Physically, this follows from
the fact that gradients of the passive scalar ($S$) are stretched and
amplified by velocity field gradients ($A_{ij}$).

The third GIM obtained with the method ($n=5$) is
\begin{equation}
	\label{eq:sigma2}
	D_{t}[ 9R T - 2Q (3U+2Q S)] = 0.
\end{equation}
This GIM is shown in Fig.~\ref{f:GCIM2}, where the solid line
indicates $ 2Q (3U+2Q S) = 9RT$, and the dashed lines again indicate 
projections of Eq.~(\ref{eq:sigma0}) on this plane. The
accumulation in this GIM becomes more clear when we restrict the DNS
data to fluid elements 
outside the manifold in Eq.~(\ref{eq:sigma0}). Again, the velocity
reduces as it goes near the GIM (see the arrows), and there is a
larger probability of finding fluid elements in the DNS in the
vicinity of these manifolds. Interestingly, the arrows align with the
GIM, and escape from zero following this manifold. This behaviour
resembles that of the Vieillefosse tail.

Finally, the system has a local central invariant manifold in the
vicinity of the fixed points. When the system is linearized around
Eq.~(\ref{eq:sigma0}), it can be shown that $U+Q S_0 = 0$ is an
invariant manifold, where $S_0$ is a fixed (constant) value of
$S$. Figure \ref{f:local_im} shows the $QS$ vs.~$U$ plane for fluid
elements with $S_0 \approx 1$, and for all fluid elements in the
DNS. The invariant manifold, and the projection of $Q = S = 0$ on this
plane are marked by solid and dashed lines respectively. Again an
accumulation of fluid elements and a slow down in the dynamics can be
seen in the DNS in the vicinity of the manifolds. Interestingly this
indicates that $S$ (i.e., the passive scalar gradients) can remain
approximately constant for a while when $S \approx -U/Q$ locally in
the flow.

{\it Discussion.} 
We presented a reduced model for the evolution of passive scalar
gradients in HIT, and a method to identify GIMs with algebraic
expressions in nonlinear ODEs. Such GIMs often arise and play a
relevant role, e.g., in reduced models for field gradients in fluids
\cite{vieillefosse_local_1982, cantwell_exact_1992,
  li_origin_2005, chevillard2006lagrangian, meneveau2011lagrangian,
  Johnson_2016, Pereira_2018, sujovolsky2019invariant}, where the
persistence of low dimensional structures in complex and
out-of-equilibrium flows is in many cases associated with the GIMs'
existence. In such models, an important problem is how to
improve approximations for the treatment of the pressure Hessian
\cite{Chevillard_2008, Wilczek_2014, Carbone_2020, Parashar_2020},
but treatment of systems with more degrees of freedom than HIT as
often appear in more realistic flows also faces the problem of how to
find invariant manifolds and to characterize phase space as the
complexity of the reduced system increases
\cite{girimaji_modified_1995, li_lagrangian_2010, pumir_2017, 
  sujovolsky2019invariant, Sujovolsky_2020}. In this case, as well as
for other ODEs with GIMs with algebraic expressions, the method
presented here can be of some value.

Moreover, the reduced model derived for the passive scalar in
Eq.~(\ref{eq:sigma0}) has two autonomous ODEs for the isotropic
invariants of the velocity field gradients ($Q$ and $R$) which are the
usual QR-model \cite{vieillefosse_local_1982,
  cantwell_exact_1992, meneveau2011lagrangian}, and three ODEs for
the evolution of the gradients of the passive scalar (the equations
for $S$, $T$, and $U$). It is remarkable that a closed system of 3
ODEs can be derived for the gradients of the scalar, which is only
limited by the approximation of neglecting diffusivity (as the pressure
Hessian does not affect the evolution of $\theta$). These ODEs and
their manifolds capture the well know physics of the problem. Passive
scalar  gradients remain the same in the absence of velocity field
gradients, and are amplified by strain in the velocity. The growth of
$S$ along these manifolds is also consistent with recent observations
of ramp-cliff structures in scalar turbulence. Other manifolds are
non-trivial as, e.g., the fact that $S$ anti-correlates with
$U/Q$, or the existence of other correlations of even higher order (in
the nonlinearity) between velocity field and scalar
gradients. Extensions of this model can be used to 
consider the problem of intermittency in the passive scalar
\cite{Celani_2000, watanabe_2007, buaria_2021}. The effect of
  diffusivity can be also included: its role should be to shrink the
  volume that orbits in phase space can explore, thus allowing studies
  of the effect of the Schmidt number in passive scalar turbulence.

{\it The authors acknowledge fruitful discussions with G.B.~Mindlin
  and D.J.~Seidler, and support from PICT Grant No.~2015-3530 and 2018-4298, and of grant UBACyT No.~20020170100508.}

\bibliography{ms}

\end{document}